\documentclass[twocolumn,amsmath,aps,nofootinbib,superscriptaddress]{revtex4-1}
 
\usepackage{amssymb}
\usepackage{amsmath}
\usepackage{epsfig}
\usepackage{subfigure}
\usepackage{mathrsfs}
\usepackage{longtable}
\usepackage{color}
\usepackage[usenames,dvipsnames]{xcolor}
\usepackage{mathtools}
\usepackage{relsize}

\usepackage{array}

\newcommand{\be}{\begin{equation}}
\newcommand{\ee}{\end{equation}}
\def\bea{\begin{eqnarray}}
\def\eea{\end{eqnarray}}

\newcommand{\we}{w_{\rm e}}

\newcommand{\rhoe}{\rho_{\rm e}}
\newcommand{\Oe}{\Omega_{\rm e}}
\newcommand{\Oemin}{\Oe^{\rm min}}
\newcommand{\Oemax}{\Oe^{\rm max}}
\newcommand{\tw}{\tilde{w}}
\newcommand{\sample}{\boldsymbol\vartheta}

\newcommand{\g}{\gamma}

\usepackage{colortbl}
\makeatletter
	
    \def\CT@@do@color{
      \global\let\CT@do@color\relax
            \@tempdima\wd\z@
            \advance\@tempdima\@tempdimb
            \advance\@tempdima\@tempdimc
    \advance\@tempdimb\tabcolsep
    \advance\@tempdimc\tabcolsep
    \advance\@tempdima2\tabcolsep
            \kern-\@tempdimb
            \leaders\vrule

                    \hskip\@tempdima\@plus  1.7fill
            \kern-\@tempdimc
            \hskip-\wd\z@ \@plus -1.7fill }
    \makeatother

\definecolor{lg}{gray}{0.85}

\begin{document}

\title{The Dark Energy Cosmic Clock: A New Way to Parametrise the Equation of State}

\author{Ewan R. M. Tarrant}
\email{ppxet@nottingham.ac.uk}

\author{Edmund J. Copeland}
\email{ed.copeland@nottingham.ac.uk}

\author{Antonio Padilla}
\email{antonio.padilla@nottingham.ac.uk}

\author{Constantinos Skordis}
\email{skordis@nottingham.ac.uk}

\affiliation{School of Physics and Astronomy, University of Nottingham, University Park, Nottingham, NG7 2RD, UK}

\date{\today}

\begin{abstract}
We propose a new parametrisation of the dark energy equation of state, which uses the dark energy density, $\Oe$ as a cosmic clock. 
We expand the equation of state in a series of orthogonal polynomials, with $\Oe$ as the expansion parameter and determine the expansion coefficients by fitting to SNIa and $H(z)$ data.
Assuming that $\Oe$ is a monotonic function of time, we show that our parametrisation performs better than the popular Chevallier--Polarski--Linder (CPL) and Gerke and Efstathiou (GE) parametrisations, and we demonstrate that it is robust to the choice of prior.
Expanding in orthogonal polynomials allows us to relate models of dark energy directly to our parametrisation, which we illustrate by placing constraints on the expansion coefficients extracted from two popular quintessence models.
Finally, we comment on how this parametrisation could be modified to accommodate high redshift data, where any non--monotonicity of $\Oe$ would need to be accounted for.
\end{abstract}

\maketitle

\section{Introduction}\label{intro}

During the last decade a vast amount of cosmological data has been collected, which indicate that a mysterious form of dark energy is driving an accelerated expansion of the Universe~\cite{Riess:2009pu,Hinshaw:2012aka,Kowalski:2008ez,Lampeitl:2009jq}. The simplest explanation for dark energy is Einstein's cosmological constant, $\Lambda$, which has a constant equation of state $\we=P_{\rm e}/\rhoe=-1$. However, $\Lambda$ suffers from several short--comings, including the fine–-tuning and coincidence problems (see e.g.~\cite{Martin:2012bt} for a review).

These thorny issues surrounding the cosmological constant have prompted investigation into alternative models such as quintessence~\cite{Caldwell:1997ii,Wetterich1988Cosmology,Ratra1988Cosmological}, Chaplygin gas~\cite{Kamenshchik:2000iv}, modified gravity~\cite{Clifton:2011jh}, holographic dark energy~\cite{Li:2004rb}, and many others, all which promote $\Lambda$ to a dynamical degree of freedom with a time--varying effective equation of state. For a review of models of dynamical dark energy see~\cite{Copeland:2006wr}. Instead of appealing to a fundamental theory to describe $\we(t)$, we can attempt to reconstruct its properties in a model independent way by proposing a functional form for $\we(t)$ and fit this directly to observation. Constraints may be placed on the dark energy equation of state once a parametrisation has been adopted. For example, by assuming a constant equation of state, $\we=w$, the authors of~\cite{Parkinson:2012vd} report $w=-1.008\pm0.085$ at $68\%$ CL, consistent with the $\Lambda$CDM concordance cosmology. Very recently, via a principle components approach, the authors of~\cite{Said:2013jxa} find that Baryon Acoustic Oscillation (BAO) data suggest deviations $w(z)<-1$ above one standard deviation at redshifts $z\sim0.25$.

Most of the $\we(t)$ parametrisations that have been proposed in the literature to date, use redshift $z$ (or equivalently scale factor $a$) as the `time' variable, i.e., $\we(t)=f(z)$. For example, the Chevallier--Polarski--Linder (CPL) parametrization, first discussed in~\cite{Chevallier:2000qy} and reintroduced in~\cite{Linder:2002et} uses a polynomial fitting function in redshift space, whilst in~\cite{Gerke:2002sx} a logarithmic expansion in $z$ was proposed. Motivated by the dynamics of quintessence models, Ref.~\cite{Corasaniti:2002vg} introduced a parametrisation dependent on five parameters which is able to reproduce the time evolution of $\we$ across a wide range of redshifts for a variety of different quintessence models. Other studies have analysed our ability to reconstruct the quintessence potential $V(\phi)$~\cite{Huterer:1998qv,Sahlen:2005zw,Li:2006ea}.

In this paper, we introduce a new parametrisation of $\we(t)$, which uses the dimensionless dark energy density fraction $\Oe(t)\equiv\rhoe(t)/3H(t)^2$ as a cosmic clock. The idea is to expand $\we(\Oe)$ in orthogonal polynomials, with $\Oe$ as the expansion parameter:
\be
\we(\Oe)=\sum_n w_n P_n(\Oe)\,.
\label{eq:the-big-idea}
\ee
This is in similar spirit to~\cite{Ferreira:2010sz} where the authors were interested in parametrising the evolution of small scale density perturbations. Parametrising $\we$ in terms of $\Oe$ was also recently explored in Ref.~\cite{Lei:2013pea}. Such an expansion has several advantages. Perhaps most importantly, $\Oe$ is a physical quantity, directly related to the properties of dark energy. Furthermore, assuming $0<\Oe<1$, it makes for an ideal expansion parameter since $\Oe$ is a naturally small number. Expanding $\we$ in terms of orthogonal polynomials has been carried out before, see for example ~\cite{Sendra:2011pt}, but with redshift as the expansion parameter.  See also Ref.~\cite{BenitezHerrera:2011wu} for examples of model independent reconstruction of the dark energy equation of state.

So long as $\Oe(t)$ remains a monotonic function over the epoch of interest it may be used as a perfectly good cosmic clock. As noted in~\cite{Johri:2000yx} it is natural to consider that $\Oe$ increases monotonically through most of cosmic history, an assumption that is well motivated by various astrophysical constraints:

\vspace{10mm}

\begin{itemize}
\item \textit{Big Bang Nucleosynthesis (BBN)}: $\Oe\lesssim 0.045$ at $z\sim10^{10}$ ~\cite{Bean:2001wt}\,,
\item \textit{Galaxy formation epoch}: $\Oe\lesssim 0.5$ at $z\sim2-4$\,,
\item \textit{Present day}: $\Oe\sim0.68$ with $\we\sim-1$ at $z=0$~\cite{Ade:2013lta}\,.
\end{itemize}

This monotonicity was recognised in~\cite{Wetterich:2004pv} and~\cite{Doran:2005sn}, where the authors proposed a dark energy parametrisation in terms of three parameters:  the dark energy equation of state today $w_0$, the amount of dark energy today $\Omega_d^0$, and the amount of dark energy at early times $\Omega_d^e$ to which it asymptotes at high redshift. Our parametrisation is similar to this in the sense that it directly relates $\we(t)$ to $\Oe(t)$, however our parametrisation does not rely on scale factor/redshift time, unlike Refs.~\cite{Wetterich:2004pv,Doran:2005sn}.

We also know that the transition from matter to dark energy domination (and hence the start of cosmic acceleration), occurred at low redshift $z\lesssim1$. This is a region that surveys such as the Dark Energy Survey (DES) will be sensitive to, and will depend crucially on $\Oe$ where the transition occurs. Our parametrisation is ideally suited to such a case since we are expanding in a small parameter around the time of transition. We can in principle use our parametrisation to reconstruct the equation of state around that region and in doing so provide another handle on understanding the nature of dynamical dark energy. 

\section{The dark energy cosmic clock}\label{sec:declock}

Assuming a Friedman--Robertson--Walker (FRW) metric, the background equations for any theory of gravity can be recast in the usual form as used in GR. The Friedman equation reads
\be
    H^2(t) \equiv \left( \frac{\dot{a}}{a} \right)^2 = \frac13\left(\rhoe + \sum_i \rho_i - \frac{3K}{a^2}\right)\,,
\label{eq:friedman}
\ee
where $a$ and $H$ are the scale factor and Hubble function respectively, $\rhoe$ is the dark energy density (which may in general be a function of additional degrees of freedom), and $\rho_i$ are the energy densities of the other possible components, including matter $\rho_{\rm m}$ and radiation $\rho_\g$. We have also allowed for a curvature term, with closed, flat and open Universes corresponding to $K=+1,0,-1$ respectively. An overdot denotes differentiation with respect to cosmic time $t$, and we use natural units throughout, with $8\pi G=m_{\rm pl}^{-2}=1$. Regardless of the theory of gravity, one may always treat dark energy as a standard fluid with a time--varying equation of state $\we(t)=P_{\rm e}(t)/\rhoe(t)$, subject to energy conservation\footnote{At the level of the background, we may always write the energy conservation equation for $\rhoe$ in this way, by absorbing any non--standard cosmological species into $\rhoe$. For example, if a fraction $f$ of cold dark matter interacts with dark energy, the contribution from $\rho_f$ may be absorbed into $\rhoe$.}: $\dot{\rho}_{\rm e}+3H(1+\we)\rhoe=0$.  Once the theory of gravity and the other components $\rho_i$ are specified, $\we$ may be subject to additional equations describing its time evolution.

\begin{figure}[!ht]
\vspace{5mm}
\includegraphics[width=8.5cm]{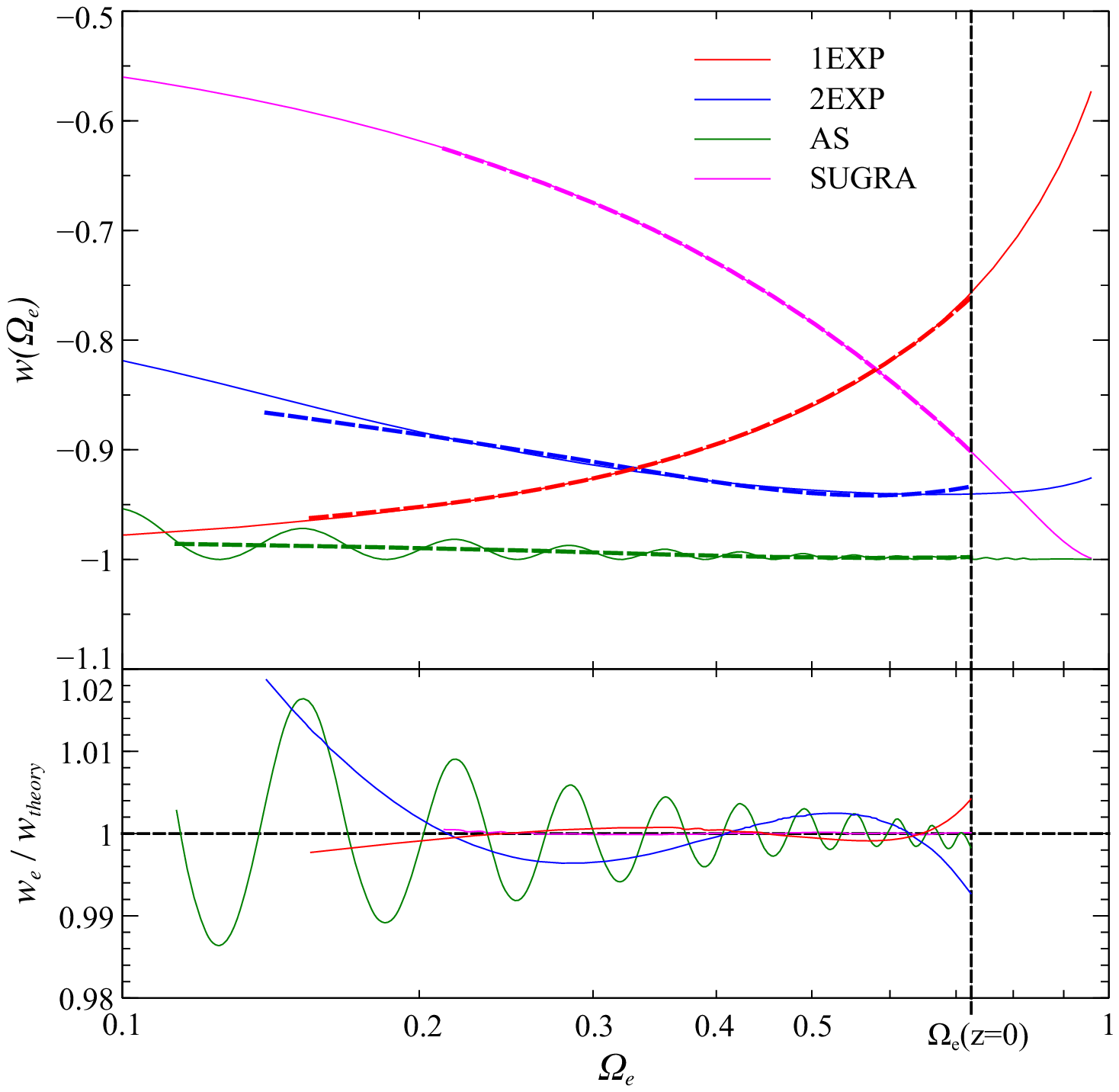} \\
\caption{The exact equation of state for four quintessence models, $w_{\rm theory}(\Oe)=P_{\rm e}/\rhoe$, (solid lines) and the corresponding reconstructed equation of state, (dashed lines) obtained by building an interpolating polynomial $\we(\Oe)=\sum_n \tw_n \tilde{U}_n(\Oe)$ over the interval $\Oe=[\Oe(z=1.75),\,\Omega_0]$, keeping the first three terms in the expansion. Time advances from left to right, and the vertical dashed line denotes $\Omega_0$, the value of $\Oe$ at $z=0$. For $\Lambda$CDM we have $\we=-1$, and so $\tilde{w}_0=-1$ and $\tilde{w}_n=0$ for all $n>0$.}
\label{fig:w-quint}
\end{figure}

The energy conservation equation for $\rhoe$ can be easily rewritten as 
\be
 \dot{\Omega}_{\rm e} = -3H \Oe \left[   \we(1 - \Omega_{\rm e})  - \sum_i w_i \Omega_i  + \frac13\Omega_K\right] \,,
\label{eq:Oedot1}
\ee
where the sum is taken over all other cosmic components $i$, and $\Omega_K\equiv-K/(aH)^2$. 

For the remainder of this paper, we will assume $\Omega_K=0$. The validity of our parametrisation, Eq.~(\ref{eq:the-big-idea}), does not rely on this assumption. Rather, setting $\Omega_K=0$ is a useful simplification when using our parametrisation, for reasons that will become clear shortly. In Section~\ref{sec:discussion} we describe how our parametrisation can be extended to accommodate $\Omega_K\neq0$. In a Universe containing dark energy, pressureless matter ($w_{\rm m}=0$), and radiation ($w_\g=\frac13$), we have $\Oe+\Omega_{\rm m}+\Omega_\g=1$, and there are turning points in $\Oe$ whenever $\we(1-\Oe)=\frac13\Omega_\g$ (assuming that $H>0$ at all times). Hence, monotonicity of $\Oe$ is broken during the radiation dominated era whenever $\we$ passes through $\we=\frac13$, and in the matter era whenever $\we$ passes through $\we=0$. At these points, the dark energy clock would `stop ticking', and our expansion Eq.~(\ref{eq:the-big-idea}) would break down. Had we included $\Omega_K$, these turning points would occur whenever $\we=\frac13(\Omega_\g-\Omega_K)/(1-\Oe)$, which clearly depends on $\Omega_K$. For small $\Omega_K$ (as is suggested by measurements of the CMB~\cite{Ade:2013lta}), the effect of including this curvature term will be  to induce a slight perturbation about the turning points $\we=\frac13$, $\we=0$ in the radiation and matter dominated eras respectively.

Many scalar field dark energy (quintessence) models possess scaling solutions on which the scalar field energy density tracks that of the dominant background fluid, $\Omega_i$. For example, in the case of a single exponential scalar field potential, $V(\phi)=V_0e^{-\lambda\phi}$, scaling solutions exist whenever $\lambda^2>3(w_i+1)$. On these solutions the scalar field dark energy equation of state mimics the evolution of the dominant background fluid, $\we=w_i$~\cite{Ferreira:1997hj,Copeland:1997et}. This is an explicit example of where we can expect the dark energy clock to break down. 

In this paper, we use our dark energy clock parametrisation to fit to low redshift background expansion data only, and so we neglect radiation. The only fixed point of Eq.~(\ref{eq:Oedot1}) that concerns us then is the matter era $\we=0$ solution. Hence, for a given set of expansion coefficients $w_n$, $\Oe$ will possess turning points whenever $\sum_n w_n P_n(\Oe)=0$. We will return to this important point in the next section. Our choice of $P_n$ is the Chebyshev polynomials of the second kind. By defining a suitable inner product we may choose the interval, $\Oe=[\Oe^{\rm min},\Oe^{\rm max}]$, over which the polynomials are orthogonal.  We denote these \textit{shifted} Chebyshev polynomials of the second kind by $\tilde{U}_n(\Oe)$, and write 
\be
\we(\Oe)=\sum_n \tw_n \tilde{U}_n(\Oe)\,.
\label{eq:we1}
\ee
The $n$ zeros (nodes) of the $\tilde{U}_n$ are useful for interpolation because the resulting interpolation polynomial minimizes Runge's phenomenon (the problem of oscillation of the interpolating polynomial near to the edges of the interval). The properties of the $\tilde{U}_n$ and how they are related to the standard Chebyshev polynomials of the second kind $U_n$, are given in Appendix~\ref{appdx:ChebyPolys}.

From the $\tilde{U}_n$ orthogonality condition, Eq.~(\ref{eq:cheby2orthoab}), we can extract the expansion coefficients $\tw_n$, given any smoothly varying monotonic $w(\Oe)$: 
\begin{widetext}
\be
\tw_n=\frac{8}{\pi(\Oe^{\rm max}-\Oe^{\rm min})^2} \int_{\Oe^{\rm min}}^{\Oe^{\rm max}} w(\Oe)\, \tilde{U}_n(\Oe) \, \sqrt{(\Oe-\Oe^{\rm min})(\Oe^{\rm max}-\Oe)}\, {\rm d}\Oe  \,.
\label{eq:cheby2coeffs}
\ee
\end{widetext}
Hence, any dark energy model which predicts a monotonic equation of state $w_{\rm theory}(\Oe)$, can be directly related to our parametrisation.  In Fig.~\ref{fig:w-quint} we show the equation of state, $w_{\rm theory}(\Oe)$, for four different quintessence potentials $V(\phi)$:
\bea
&{\rm 1EXP}& \quad  V_0e^{\lambda\phi}\,\, \text{ \cite{Ferreira:1997hj,Copeland:1997et}}\,, \nonumber \\
&{\rm 2EXP}& \quad  V_0[e^{\lambda_1\phi}+e^{\lambda_2\phi}]\,\, \text{\cite{Barreiro:1999zs}}\,, \nonumber\\
&{\rm AS}&     \quad  V_0\left[\lambda_1 + (\lambda_2 - \phi)^2  \right]e^{-\lambda_3\phi}\,\, \text{\cite{Albrecht:1999rm,Skordis:2000dz}}\,, \nonumber \\
&{\rm SUGRA}&     \quad  V_0^{4+\lambda}/\phi^{\lambda}e^{\phi^2/2}\,\, \text{\cite{Brax:1999gp}}\,. \nonumber
\eea
In the same figure, we also plot $\we(\Oe)$, the `reconstructed' equation of state built from shifted Chebyshev polynomials orthogonal on the interval $\Oe=[\Oe^{\rm min},\Oe^{\rm max}]$, where $\Oe^{\rm min}=\Oe(z=1.75)$ and $\Oe^{\rm max}=\Omega_0=\Oe(z=0)$. In this section we shall only work up to second order in expansion, for reasons that will be made clear shortly. The $\tw_n$ as given by Eq.~(\ref{eq:cheby2coeffs}) depend on the upper and lower limits of integration $\Oe^{\rm max}$ and $\Oe^{\rm min}$.  When fitting to a known model (such as 1EXP or SUGRA), these limits are unambiguously defined: we can for example adjust the height of the potential (set by $V_0$) to give $\Oe=0.72$ for each model, whilst we can easily compute the lower limit $\Oe(z=1.75)$ numerically. This is the procedure which was followed for the models of Fig.~\ref{fig:w-quint}.  When fitting to observational data however -- where the model is \emph{not} known, these limits must be treated with care, and this is something we return to discuss in Section~\ref{sec:theory}. We also point out that since the dynamics of the scalar field is different for each quintessence model in Fig.~\ref{fig:w-quint}, the evolution of the dark energy density is also different, and so the lower limit $\Oe(z=1.75)$, will in general be different from model to model.\footnote{A cosmological constant ($\we=-1)$, maximally decreases $\Oe(z)$ with increasing redshift relative to the non--phantom quintessence models considered here. For $\Lambda$CDM, with $\Omega_{\Lambda}(z=0)=0.7$, we have $\Omega_{\Lambda}$=0.1 at $z\simeq1.75$.}

As can be seen from the  Fig.~\ref{fig:w-quint}, our expansion does well in capturing the evolution of $\we$. The rapid oscillations seen for the AS model are induced by the late time oscillations of the field about its minimum. Our parametrisation is unable to resolve the individual oscillations (to do so would require retaining a large number of terms in the expansion), but it does capture the average behaviour rather well. Furthermore, for all four models, the series rapidly converges. For example for the 1EXP model we find: $\tw_0=-0.876$, $\tw_1=0.050$, $\tw_2=0.005$. In general of course, the rate of convergence will depend upon the behaviour of the function $\we(\Oe)$ that we are trying to reconstruct. If, as various astrophysical constraints suggest, $\Oe$ becomes less important at high redshift, our expansion should in principle always converge, as higher order terms $\Oe^n$ for large $n$ will become negligible. Of course when confronted with data, the order at which the equation of state expansion is truncated depends upon the quality of the available data.\\

Having motivated our $\we$ parametrisation, we now turn our attention to assessing its performance by fitting to background expansion data. To aid this analysis we first present analytic solutions to the background equations of motion for dark energy which are valid up to second order in the expansion of $\we(\Oe)$.\\

\textit{Scale factor, $a(\Oe)$}:  We begin with our equation of state parametrisation in terms of Chebyshev polynomials, Eq.~(\ref{eq:we1}), which may be rewritten at second order as:
\be
\we(\Oe)=w_0+w_1\Oe+w_2\Oe^2\,.
\label{eq:we2ndorder}
\ee
The coefficients, $w_0$, $w_1$ and $w_2$ are combinations of the $\tw_n$ and are given in Appendix~\ref{appdx:ChebyPolys}. Neglecting radiation, Eq.~(\ref{eq:Oedot1}) may be rearranged to give:
\be
-3\int^{a}_{a_0}\frac{{\rm d}a}{a}=\int^{\Oe}_{\Omega_0}\frac{{\rm d}\Oe}{\Oe(1-\Oe)(w_0+w_1\Oe+w_2\Oe^2)}\,,
\label{eq:scalefactorintegral}
\ee
where the lower limits of the integrals (subscript $0$) denote the value today, and we have written $\Oe(z=0)=\Omega_0$. We remind the reader that $w_0$ is not the value of $\we$ today, but is the zeroth order expansion parameter. The LHS of Eq.~(\ref{eq:scalefactorintegral}) is trivial, and the RHS may be expanded using partial fractions and the resulting terms integrated separately. We find:
\be
a^{-3}(\Oe)=\left(\frac{\Oe}{\Omega_0}\right)^{\alpha-1}\left(\frac{1-\Oe}{1-\Omega_0}\right)^{\beta+1} \left(\frac{\we(\Oe)} {w(\Omega_0)}\right)^\g \,e^{F(\Oe)}\,,
\label{eq:scalefactorsol}
\ee
where we have set $a_0=1$. The powers $\alpha$, $\beta$ and $\g$ are combinations of the $w_n$, and correspond to coefficients of the partial fraction expansion of Eq.~(\ref{eq:scalefactorintegral}). As a result, if any one or more of the $\we(\Oe)$ expansion coefficients $w_0$, $w_1$ or $w_2$ in Eq.~(\ref{eq:scalefactorintegral}) is exactly zero, then the powers $\alpha$, $\beta$ and $\g$ and the function $F(\Oe)$ will change. In the case where none of the $w_n$ are zero:
\begin{eqnarray}
\alpha &=& \frac{1+w_0}{w_0}\,, \quad \beta = -\frac{1+w_T}{w_T}\,, \nonumber \\
\g &=& -\frac{1}{2w_0}\frac{w_1+w_2}{w_T}\,, \quad  w_T = w_0+w_1+w_2\,,
\end{eqnarray}
and 
\be
\begin{split}
F(\Oe) = & \frac{1}{w_0w_Tq} \left[w_1(w_1+w_2)-2w_0w_2\right]  \\
	     & \times {\rm arctan}\left[\frac{2w_2q(\Omega_0-\Oe)}{(2w_2\Omega_0+w_1)(2w_2\Oe+w_1)+q^2}\right]\,,
\end{split}
\ee
where $q=\sqrt{4w_0w_2-w_1^2}$. The solutions corresponding to the six different cases of zero $w_0$, $w_1$ or $w_2$ are listed in Table~\ref{tab:particularsols} of Appendix~\ref{appdx:particularsols}. Notice that the solution~(\ref{eq:scalefactorsol}) breaks down at $\Oe=0$, $\Oe=1$ and $\we(\Oe)=w_0+w_1\Oe+w_2\Oe^2=0$. These are fixed points of Eq.~(\ref{eq:Oedot1}), and reflect the fact that at these points $\Oe$ cannot be used to measure time.  \\

\textit{Dark energy density $\rhoe(\Oe)$}: The perfect fluid equation of motion for dark energy reads:
\be
\dot{\rho}_{\rm de}=-3H\left[1+\we(\Oe)\right]\rhoe\,.
\label{eq:rhoe}
\ee
Using Eqs.~(\ref{eq:we2ndorder}) and~(\ref{eq:Oedot1}) in the above equation and dropping radiation we have that:
\be
\int^{\rhoe}_{\rho_0}\frac{{\rm d}\rhoe}{\rhoe}=\int^{\Oe}_{\Omega_0}\frac{{\rm d}\Oe}{\Oe(1-\Oe)} \left[ 1+ \frac{1}{(w_0+w_1\Oe+w_2\Oe^2)}\right]\,.
\label{eq:rhoeintegral}
\ee
The integrals are the same as those that were required for the $a(\Oe)$ solution, and we find:
\be
\rhoe(\Oe)=\rho_0\,\left(\frac{\Oe}{\Omega_0}\right)^\alpha\left(\frac{1-\Oe}{1-\Omega_0}\right)^\beta\left(\frac{\we(\Oe)} {\we(\Omega_0)}\right)^\g \,e^{F(\Oe)}\,.
\label{eq:rhoesol}
\ee
%

\textit{Hubble rate, $H(\Oe)$}: The Hubble rate is simply
\be
H(\Oe)=\sqrt{\frac{\rhoe(\Oe)}{3\Oe}}\,,
\label{eq:hubblesol}
\ee
with $\rhoe(\Oe)$ given by equation Eq.~(\ref{eq:rhoesol})\\

\textit{Angular diameter distance, $d_A(\Oe)$}: In a flat Universe, the angular diameter distance to an object at redshift $z=(1/a-1)$, is given by:
\be
d_A(z)=(1+z)^{-1}\int^{t_0}_{t}\frac{{\rm d}t}{a} \,.
\label{eq:angulardiamdist}
\ee
Again neglecting radiation, we can use Eq.~(\ref{eq:Oedot1}), to substitute for ${\rm d}t$ to give
\be
\begin{split}
d_A(\Oe) &=a(\Oe)\,G_0 \, \times \\
& \int^{\Oe}_{\Omega_0}{\rm d}\Oe\,  \Oe^{\bar{\alpha}} (1-\Oe)^{\bar{\beta}} (w_0+w_1\Oe+w_2\Oe^2)^{\bar{\g}} \,e^{\bar{F}}\,,
\label{eq:angulardiamdistintegral}
\end{split}
\ee
where $\bar{\alpha}=-\frac16(\alpha+5)$, $\bar{\beta}=-\frac16(\beta+4)$, $\bar{\g}=-\frac16(\g+6)$ and $\bar{F}=-\frac16F$. The constant $G_0=\sqrt{\frac{1}{3\rho_0}}\Omega_0^{\alpha_0}(1-\Omega_0)^{\beta_0}(w_0+w_1\Omega_0+w_2\Omega_0^2)^{\gamma_0}$, where $\alpha_0=\frac16(\alpha+2)$, $\beta_0=\frac16(\beta-2)$, and $\g_0=\frac16\g$. This integral can be performed analytically if $w_2=0$:
\be
\begin{split}
d_A(\Oe) & =a(\Oe)\,G(\Oe) \, \times \\
& F_1\left(\bar{\alpha}+1;\, -\bar{\beta}\,,-\bar{\g};\, \bar{\alpha}+2;\,\Oe\,, -\frac{w_1}{w_0}\Oe  \right) + {\rm const.}
\label{eq:comvingdistSol}
\end{split}
\ee
where $F_1(...)$ is the Appell hypergeometric function~\cite{Abramowitz1965Handbook}, and 
\be
G(\Oe)=\frac{G_0}{\bar{\alpha}+1}\Oe^{\bar{\alpha}+1}w_0^{\bar{\g}}\,.
\label{eq:comvingdistSolG}
\ee
Up to a constant, this is the final result for the angular diameter distance to first order in the expansion of $\we(\Oe)$.

\section{Constraints from observational data}\label{sec:data}

In this section we present constraints on the parameters of our dark energy equation of state parametrisation Eq.~(\ref{eq:we1}), by performing a global Monte Carlo Markov Chain (MCMC) fit to data. We take the set of base parameters
\be
\sample  = \{ \Omega_mh^2,\, H_0,\,  \tw_n \} \,,
\label{eq:baseParams}
\ee
and use a modified version of the \texttt{CosmoMC} code~\cite{Lewis:2002ah} to sample from the joint posterior distribution of these parameters,
\be
\mathcal{P}(\boldsymbol\vartheta | \boldsymbol x) = \frac{\mathcal{L}(\boldsymbol x | \boldsymbol\vartheta)\mathcal{P}(\boldsymbol\vartheta)} 
    {\int {\rm d}\boldsymbol\vartheta\, \mathcal{L}(\boldsymbol x | \boldsymbol\vartheta)\mathcal{P}(\boldsymbol\vartheta)} \,,
\label{eq:posterior}	
\ee
where $\mathcal{L}(\boldsymbol x | \boldsymbol\vartheta)$ is the likelihood of the data $\boldsymbol x$ given the model parameters $\boldsymbol\vartheta$ and $\mathcal{P}(\boldsymbol\vartheta)$ is the prior probability density. Other parameters, such as $\Omega_0=\Oe(z=0)$ may be derived from this base set. We refer to a single realization of $\boldsymbol\vartheta$, as a \textit{sample}. 

The data we use are the compilation of differential--age measurements of the Hubble rate $H(z)$~\cite{Moresco:2012by}, the latest and most precise (local) estimate of the Hubble constant $H_0$~\cite{Riess:2011yx}, and the Union2 SNIa compilation~\cite{Amanullah:2010vv}. These data span a redshift range $z=0\, -\,1.75$.\\

Our parametrisation is only valid if $\Oe$ is a monotonic function over the redshift range of interest. Monotonicity of $\Oe$ is broken whenever $\we(\Oe)=\sum_n \tw_n \tilde{U}_n(\Oe)=0$. Since we know that $\we(z=0)$ is well constrained to be negative by a variety of different observations~\cite{Riess:2009pu,Hinshaw:2012aka,Kowalski:2008ez,Lamastra:2011sq}, we impose the hard prior $\we(\Oe)<0$ at \textit{all} redshifts of interest, $z=0\, -\,1.75$. This corresponds to\footnote{We remind the reader that this prior would be modified if $\Omega_K\neq0$.} $\dot{\Omega}_{\rm e}>0$.  So, for a given sample $\sample$, if any single data point within this redshift range yields $\we(\Oe)\ge0$ then the entire sample is rejected and does not feature in the evaluation of the likelihood function. For monotonicity to be broken within the interval $z=0\, -\,1.75$, a fairly rapidly varying equation of state would be required, which is disfavoured given existing constraints~\cite{Said:2013jxa}.

If we truncate the expansion of $\we(\Oe)$ at second order, we can take advantage of our analytic solutions of Section~\ref{sec:declock} when numerically implementing this prior. Firstly, the roots of $\we(\Oe)=w_0+w_1\Oe+w_2\Oe^2$ are computed:
\be
\Omega_{\pm}=\frac{-w_1\pm\sqrt{w_1^2-4w_0w_2}}{2w_0}\,.
\label{eq:wroots}
\ee
We remind the reader that the $w_n$ are related to $\tw_n$ through Eq.~(\ref{eq:wtowtilde}). The prior on $\we$ today becomes $w_0+w_1\Omega_0+w_2\Omega_0^2<0$. If $w_1^2>4w_0w_1$, then depending on the values of $w_0$, $w_1$ and $w_2$, there are three regions of monotonic $\Oe$ that can be defined. These are summarised (along with the case of $w_2=0$) in Table~\ref{tab:montonicintervals}.
\renewcommand*\arraystretch{1.2}
\begin{table}[t]
\hfill{}
\begin{tabular}{c|c}
\hline
\hline
   condition     								&  monotonic interval \\
\hline
  $w_2>0$,  $\Omega_-<\Omega_0<\Omega_+$	        &   $[{\rm max}(\Omega_-,0),{\rm min}(\Omega_+,1)]$\\
  $w_2<0$,  $\Omega_+<\Omega_0$	    			&   $[{\rm max}(\Omega_+,0),1]$\\
  $w_2<0$,  $\Omega_->\Omega_0$	    			&   $[0,{\rm min}(\Omega_-,1)]$ \\
\hline
  $w_2=0$, $w_1>0$        						&   $[0,{\rm min}(\Omega_-,1)]$ \\
   $w_2=0$, $w_1<0$        						&   $[{\rm max}(\Omega_-,0),1]$
\end{tabular}
\hfill{}
\caption{The intervals where $\dot{\Omega}_{\rm e}>0$ (if radiation can be neglected) for $\we(\Oe)=w_0+w_1\Oe+w_2\Oe^2$.}
\label{tab:montonicintervals}
\end{table}
Alternatively, if $w_1^2<4w_0w_1$ then there are no real roots, and $\we(z=0)<0$ is sufficient to guarantee that $\Oe$ remains monotonic at all times.

We would like to convert the redshift $z_i$ (or scale factor $a_i$) of each data point to a density $\Omega_{{\rm e}_i}$. Since Eq.~(\ref{eq:scalefactorsol}) cannot be inverted analytically, this must be done numerically. We use a simple bisection method, where the boundaries of the search interval over $\Oe$ correspond to the limits of the monotonic region of $\Oe$ given in Table~\ref{tab:montonicintervals}. Since $\Oe$ is by definition monotonic in this region, there will only be one single root, $\Oe$, to the equation $a(\Omega_{{\rm e}_i})-a_i=0$. If the bisection fails to find a solution in this interval, then the solution does not exist in this interval, and must exist where $\we\ge0$. If this is the case, the sample is rejected, and does not feature in the evaluation of the likelihood function. Those samples which generate $\we<0$ for $z\le1.75$ but $\we>0$ for $z>1.75$ are still accepted, since our demand is that $\Oe$ need only be monotonic over the region where our data lie, $z\le1.75$.

We emphasise that the mapping from $a_i$ to $\Omega_{{\rm e}_i}$ does not rely on having an analytic solution $a(\Oe)$. The background equations can be easily solved numerically, and so the mapping from $a_i$ to $\Omega_{{\rm e}_i}$ is trivial once the regions of monotonic $\Oe$ are known. Hence, in principle the $a_i\rightarrow\Omega_{{\rm e}_i}$ mapping can be performed for an \textit{arbitrary} number of terms in the Chebyshev expansion of $\we$. 

The data we use does not constrain the expansion coefficients of order 2 or higher, and hence we truncate our expansion at first order. Hence, we retain only the first two terms in the expansion, and so specialise to the case:
\be
\sample  = \{ \Omega_mh^2,\, H_0,\,  w_0,\, w_1 \} \,.
\label{eq:baseParams2}
\ee
Now we summarise the data that we use in our MCMC analysis.\\

\textit{HST $H_0$ prior}: The current best (local) measurement of the Hubble constant comes from the observation of Type Ia supernovae (SNe Ia) via the Wide Field Camera 3 (WFC3) on the Hubble Space Telescope (HST). This estimate is $H_0=(73.8\pm2.4)\,{\rm km\,s}^{-1}\,{\rm Mpc}^{-1}$ which includes systematic errors, corresponding to a $3.3\%$ uncertainty~\cite{Riess:2011yx}\footnote{This local measurement of $H_0$ is in tension with the recent Planck measurement~\cite{Ade:2013lta}, $H_0=(67.3\pm1.2)\,{\rm km\,s}^{-1}\,{\rm Mpc}^{-1}$, from CMB data alone. For a discussion of the differences between these measurements see~\cite{Ade:2013lta}.}. \\

\textit{$H(z)$ from differential--age techniques}: A weakness of supernova observations, BAO angular clustering, weak lensing, and cluster--based measurements is that they rely on an integral of the expansion history, rather than the expansion history itself. The differential--age technique circumvents this limitation by measuring the integrand ${\rm d}z/{\rm d}t_e$ directly, or in other words, the \textit{change} in the age of the Universe as a function of redshift. This can be achieved by measuring the ages of passively evolving galaxies with respect to a fiducial model, and so does not rely on computing absolute ages. We use the compilation of eighteen measurements of Hubble rate $H_{\rm obs}(z)$ that are quoted in~\cite{Moresco:2012by}, which span the redshift range $z=0\, -\,1.75$. For each of the eighteen data points $i$, we convert from redshift $z_i$, to dark energy density $\Omega_{{\rm e}_i}$ via the bisection algorithm discussed above, and fit our samples by minimising:
\be
\chi^2_{H(z)} = \sum_{i}\frac{\left[ H(\Omega_{{\rm e}_i},\sample) - H_{\rm obs}(\Omega_{{\rm e}_i}) \right]^2}{2\sigma_i^2(\Omega_{{\rm e}_i})}\,,
\label{eq:SNIachi2}
\ee
where $\sigma_i$ are the measurement variances. We use Eq.~(\ref{eq:hubblesol}) to compute $H(\Omega_{{\rm e}_i},\sample)$.\\

\textit{Union2 SNIa sample}: We use the Union2 SNIa compilation released by the Supernova Cosmology Project (SCP)~\cite{Amanullah:2010vv}, which consists of 557 data points, spanning a redshift range $z=0\,-\,1.4$. The statistical analysis of such SN samples rests on the definition of the distance modulus:
\be
\mu(z_i)=5{\rm log}_{10}[(1+z_i)^2d_A(z_i,\sample)]+25 + \mu_0\,,
\label{eq:SNIadistmodulus}
\ee
where the angular diameter distance $d_A(z_i,\sample)$ was defined in Eq.~(\ref{eq:angulardiamdist}). The nuisance parameter $\mu_0$, encodes the value of $H_0$, over which we analytically marginalise with a flat prior. This is the standard marginalisation procedure, see for example~\cite{Lewis:2002ah,Bridle:2001zv}, and is equivalent to marginalising over SNIa absolute magnitude. We convert from $z_i$ to $\Omega_{{\rm e}_i}$, and minimise the following expression:
\be
\chi^2_{\rm SNIa} = {\bf d}^T {\bf C}^{-1}_{\rm SNIa} {\bf d} - \frac{ {\bf d}^T ({\bf C}^{-1}_{\rm SNIa})^2 {\bf d} }{{\bf C}^{-1}_{\rm SNIa}}\,.
\label{eq:SNIachi2}
\ee
The column vector ${\bf d}$ contains the theoretical minus observed distance moduli: 
\renewcommand*\arraystretch{1.9}
\begin{center}
\begin{table*}[t]
\hfill{}
\begin{tabular}{c|c|c|c|c}
\hline
\hline
                                                      &  \multicolumn{2}{c|}{Dark Energy Clock}                                      &      CPL                                       &      GE   \\
\hline
     	   		  			&  flat priors on $w_0$ and $w_1$ &  flat priors on $\we|_{z=0}$ and $\we^{\prime}|_{z=0}$ 	  &   	    &     \\   
\hline
    $\Omega_{\rm m}h^2$   		&      $0.127^{+0.04}_{-0.03} $     &    $0.168^{+0.04}_{-0.04} $      &    $0.188^{+0.03}_{-0.03}$      &      $0.175^{+0.02}_{-0.02}$ \\
    $H_0$    				&      $72.13^{+1.76}_{-1.75} $     &    $71.89^{+1.80}_{-1.80} $      &    $71.93^{+1.78}_{-1.77}$      &      $71.84^{+1.77}_{-1.76}$ \\
    $w_0$  					&      $0.24^{+2.44}_{-2.29}$        &    $-1.53^{+1.49}_{-1.63} $      &    $-1.14^{+0.27}_{-0.28}$      &      $-1.14^{+0.26}_{-0.27}$ \\    
    $w_1$  					&      $-1.45^{+2.74}_{-2.80}$      &    $0.74^{+2.61}_{-2.49} $        &    $-1.64^{+2.76}_{-2.90}$      &      $1.78^{+2.45}_{-2.38}$ \\
\hline
    $\Omega_0$  				&      $0.76^{+0.06}_{-0.07}$       &\cellcolor{lg}$0.67^{+0.06}_{-0.06}$      &    \cellcolor{lg}$0.674^{+0.07}_{-0.07}$      &      \cellcolor{lg}$0.660^{+0.06}_{-0.06}$\,\, \\
    $\we|_{z=0}$  			&      $-1.15^{+0.24}_{-0.24}$     &     \cellcolor{lg}$-1.21^{+0.25}_{-0.25}$       &    \cellcolor{lg}$-1.14^{+0.27}_{-0.28}$      &      \cellcolor{lg}$-1.14^{+0.26}_{-0.27}$ \\    
    $\we^{\prime}|_{z=0}$  		&      $0.37^{+1.2}_{-1.4}$           &     \cellcolor{lg}$-0.95^{+1.70}_{-1.79}$       &    \cellcolor{lg}$-1.64^{+2.76}_{-2.90}$      &      \cellcolor{lg}$-1.78^{+2.38}_{-2.45}$ \\
\hline
   $\chi^2_{\rm dof}$      		&      $0.945$                        	&    $0.936$                         	      &     $0.945$                           	    &      $0.945$ \\

\end{tabular}
\hfill{}
\caption{Maximum likelihood parameter values and their 68\% CL upper and lower limits. The top third of the table gives constraints on the base parameters, whilst the middle third gives constraints on the derived parameters. The lower third gives the $\chi^2$ per degree of freedom. 
 For our Dark Energy Clock parametrisation, we show constraints on the expansion parameters assuming flat priors on $w_0$ and $w_1$ (so \textit{non--flat} priors on $\we|_{z=0}$ and $\we^{\prime}|_{z=0}$) and constraints with flat priors on $\we|_{z=0}$ and $\we^{\prime}|_{z=0}$. The shaded region indicates the constraints which may be directly compared between three parametrisations.}
\label{tab:results}
\end{table*}
\end{center}
\be
d_i=5{\rm log}_{10}[d_A(\Omega_{{\rm e}_i},\sample)a^{-2}(\Omega_{{\rm e}_i},\sample)]+25 - \mu_{\rm obs}(\Omega_{{\rm e}_i})\,,
\label{eq:dvector}
\ee
whilst ${\bf C}_{\rm SNIa}=C_{\rm SNIa}(\Omega_{{\rm e}_i},\Omega_{{\rm e}_j})$ is the covariance matrix. We use Eq.~(\ref{eq:scalefactorsol}) to compute $a^{-2}(\Omega_{{\rm e}_i},\sample)$ and perform the integral for $d_A(\Omega_{{\rm e}_i},\sample)$ in Eq.~(\ref{eq:angulardiamdistintegral}) numerically (with $w_2=0$). The total $\chi^2$ to be minimised is then
\be
\chi^2 = \chi^2_{H_0} + \chi^2_{H(z)} + \chi^2_{\rm SNIa} \,.
\label{eq:chi2}
\ee

Taking the base set of parameters~(\ref{eq:baseParams2}), we compare the constraints obtained on the expansion parameters of our $\we(\Oe)$ parametrisation 
\be
\we(\Oe)= w_0+w_1\Oe\,,
\label{eq:TSPC}
\ee
against two common parametrisations that may be found in the literature: the Chevallier--Polarski--Linder (CPL) parametrisation~\cite{Chevallier:2000qy,Linder:2002et}, which is currently favoured by the WMAP team~\cite{Hinshaw:2012aka}:
\be
\we^{\rm CPL}(z)= w^{\rm CPL}_0 + w^{\rm CPL}_1\left[\frac{z}{1+z}\right]\,,
\label{eq:CPL}
\ee
and the parametrisation of Gerke and Efstathiou (GE)~\cite{Gerke:2002sx}:
\be
\we^{\rm GE}(z)=w^{\rm GE}_0 + w^{\rm GE}_1\left[ {\rm ln}\,\left(\frac{1}{1+z}\right) \right]\,.
\label{eq:GE}
\ee
We begin by assuming flat, uninformative priors on all base parameters, ensuring that they are wide enough such that they do not affect the posterior distributions of the parameters. For all three parametrisations, these prior ranges are as follows: $\Omega_{\rm m}h^2\in[0.01,0.99]$, $H_0\in[50,90]$, $w_0\in[-12,12]$ and $w_1\in[-20,20]$. We also impose the hard prior $\we(z)<0$ for $z\le1.75$ for the CPL and GE parametrisations in order to facilitate a fair comparison to our parametrisation which \textit{requires} this prior in order to be valid. The maximum likelihood values of the base parameters and their $1-\sigma$ deviations are summarised in the top third of Table~\ref{tab:results}.  

Since we allow $\Omega_0$ (the dark energy density today) to vary, our equation of state parametrisation, Eq.~(\ref{eq:TSPC}), actually has \textit{three} free parameters, $w_0\,,w_1$ and $\Omega_0$. This is different to CPL and GE, which depend on only $w_0$ and $w_1$. In order to directly compare the three parametrisations then, we also show constraints on the derivatives of $\we$ at $z=0$. For the CPL parametrisation we have $\we^{\rm CPL} |_{z=0}=w_0^{\rm CPL}$ and ${\rm d}\we^{\rm CPL} / {\rm d}z |_{z=0}=w_1^{\rm CPL}$, whilst for the parametrisation of Gerke and Efstathiou, we have $\we^{\rm GE} |_{z=0}=w_0^{\rm GE}$ and ${\rm d}\we^{\rm GE} / {\rm d}z |_{z=0}=-w_1^{\rm GE}$. The equivalent expansion parameters in our parametrisation are given by:
\bea
\we|_{z=0} &=& w_0 + w_1\Omega_0\,,  \nonumber \\
\frac{{\rm d}\we}{{\rm d}z}\Big|_{z=0} &=& 3w_1(w_0 + w_1\Omega_0)\Omega_0 (1-\Omega_0)\,.
\label{eq:DECC-derived}
\eea
From these two equations, it is clear how our parametrisation depends upon $\Omega_0$. This comparison in terms of derivatives of $\we$ is necessary if we wish to compare like--for--like expansion parameters. 

\begin{figure*}[t]
\vspace{-30mm}
	\begin{tabular}{cc}
		\includegraphics[width=9.2cm]{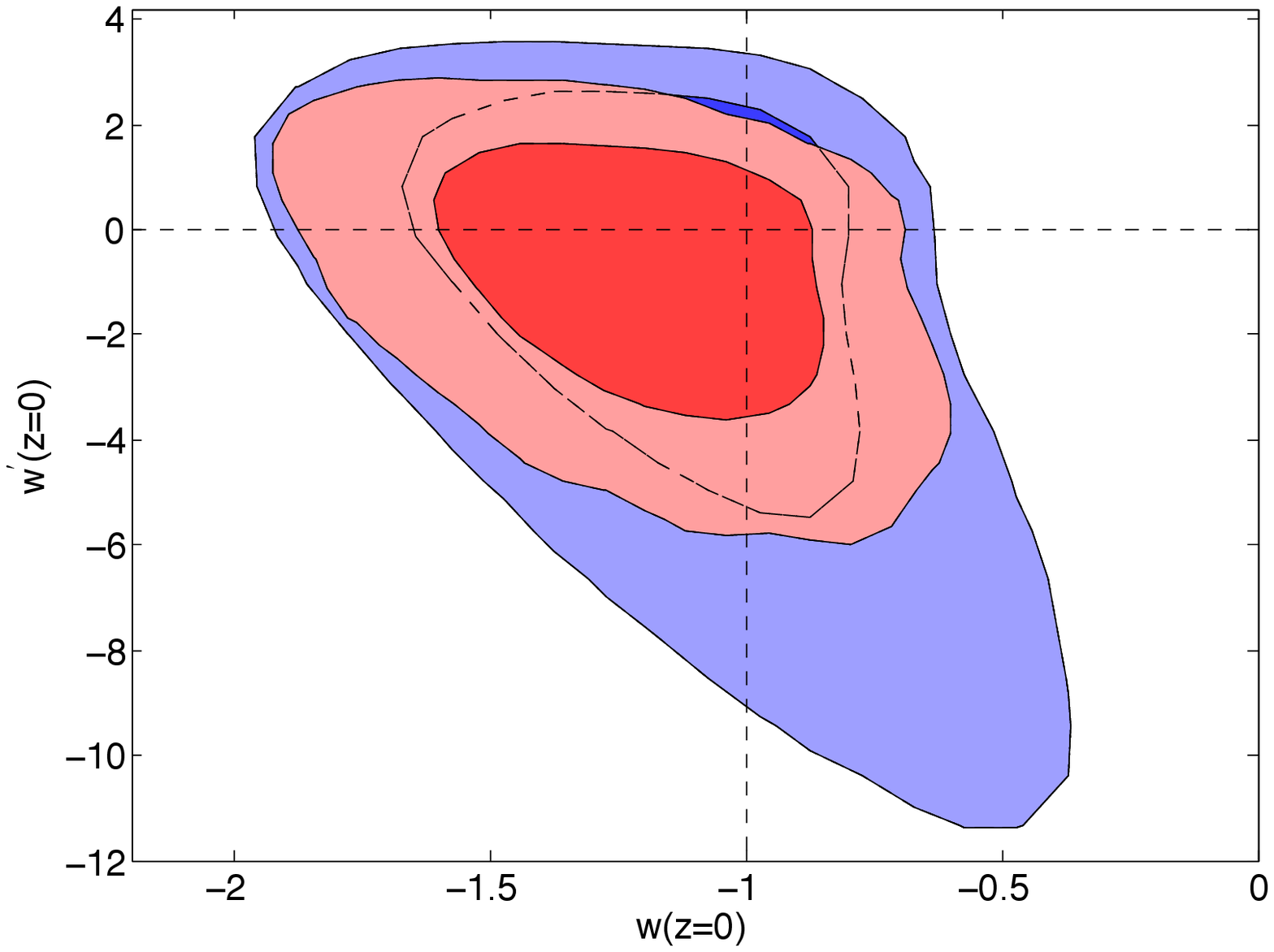} &
		\includegraphics[width=9.2cm]{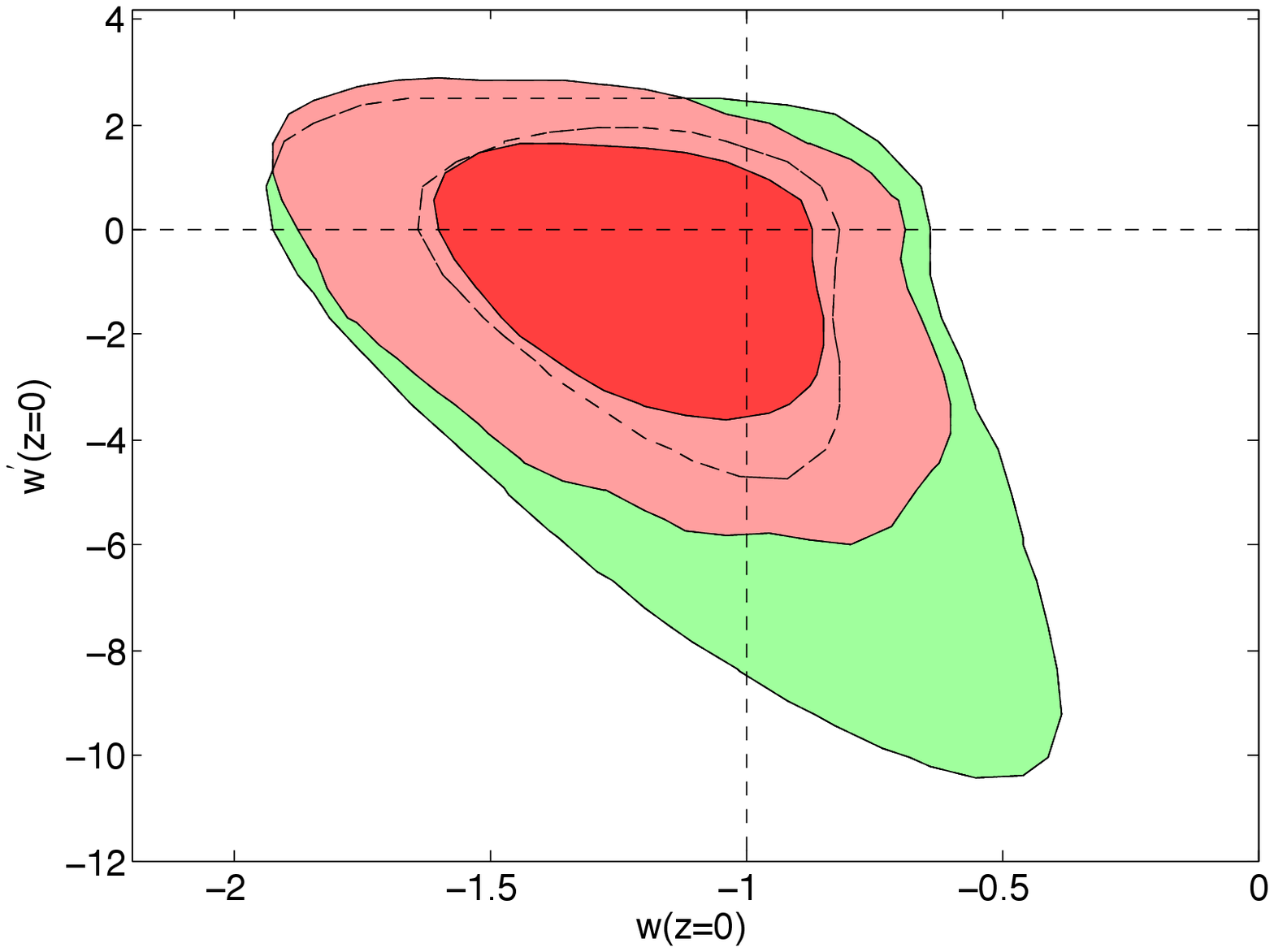} \\
	\end{tabular}
\vspace{-30mm}
	\caption{2D 68\% (dark shading) and 95\% (light shading) marginalised contours in the $\we|_{z=0}$--$\we^\prime|_{z=0}$ plane. \textit{Left panel}: Red regions: DE--clock  with flat priors on $\we|_{z=0}$ and $\we^\prime|_{z=0}$; blue regions: CPL parametrisation.   \textit{Right panel}: Red regions: DE--clock with flat priors on $\we|_{z=0}$ and $\we^\prime|_{z=0}$; green regions: GE parametrisation.   }
	\label{fig:2D_w-wprime}
\end{figure*}

Now, since we have assumed flat priors on all of our base parameters, the derived parameters $\we^{\rm CPL, GE} |_{z=0}$ and ${\rm d}\we^{\rm CPL, GE} / {\rm d}z |_{z=0}$ will also have flat priors, since these derived parameters are simply linear combinations of the base parameters. For our dark energy cosmic clock parametrisation however, the analogous derived parameters, Eq.~(\ref{eq:DECC-derived}), will \textit{not} have flat priors since they are non--linear combinations of the base parameters. In order to check that our derived parameters, $\we|_{z=0}$ and $\we^\prime|_{z=0}={\rm d}\we/{\rm d}z|_{(z=0)}$ are robust to the choice of prior, we adjust the sample likelihoods and weights in our MCMC chains for the dark energy clock parametrisation in order to obtain flat priors on Eq.~(\ref{eq:DECC-derived}):
\be
-\rm{ln}\,\mathcal{L} \rightarrow -\rm{ln}\,\mathcal{L}  + \rm{ln}\,|{\bf J}|\,, \quad \quad  {\rm weight}\rightarrow{\rm weight}\times|{\bf J}|\,.
\label{eq:flatten-priors}
\ee
Here $|{\rm\bf J}|=|det({\rm J})|=|\,3(w_0+w_1\Omega_0^2)(1-\Omega_0)\,|$. The constraints on $\we|_{z=0}$ and $\we^{\prime}|_{z=0}$ with both flat and non--flat priors are shown in the middle third of Table~\ref{tab:results}. As can be seen from the Table, the constraints on $\we|_{z=0}$ are robust to changing the prior, whilst $\we^{\prime}|_{z=0}$ shows a weak dependence on the prior. To compare the expansion parameters between the three parametrisation, one must consider the constraints given in the shaded region of Table~\ref{tab:results}. This region displays one of the main results of our paper, which demonstrates that we obtain tighter constraints on the dark energy equation of state using our parametrisation, compared to CPL and GE.  In Fig.~\ref{fig:2D_w-wprime} we compare the 2D $68\%$ and $95\%$ marginalised contours in the $\we|_{z=0}$--$\we^\prime|_{z=0}$ plane between the three parametrisations under consideration. All three parametrisations depicted in Fig.~\ref{fig:2D_w-wprime} have flat priors on $\we|_{z=0}$ and $\we^\prime|_{z=0}$. 

\begin{figure*}[t]
\vspace{-30mm}
	\begin{tabular}{cc}
		\includegraphics[width=9.2cm]{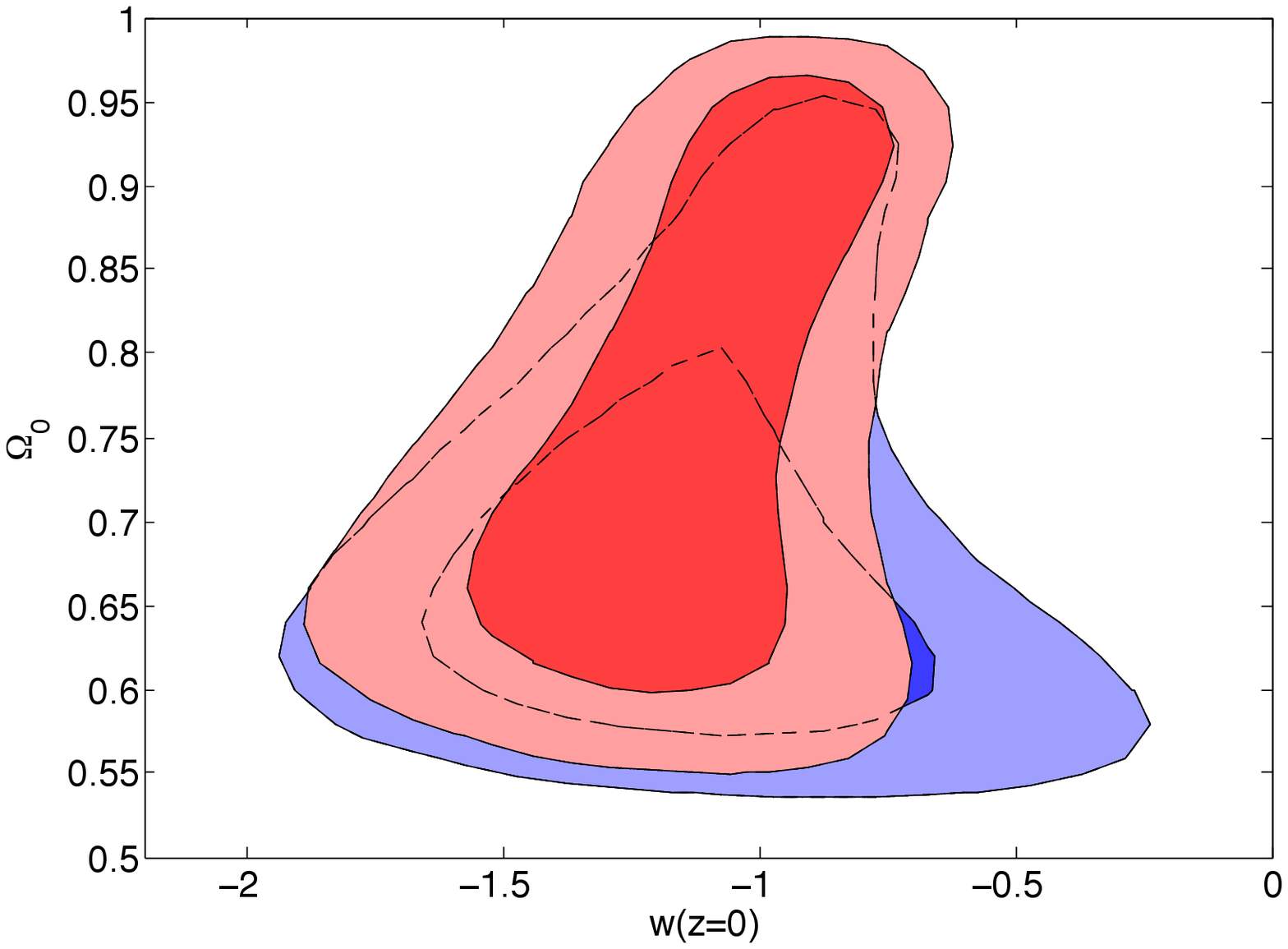} &
		\includegraphics[width=9.2cm]{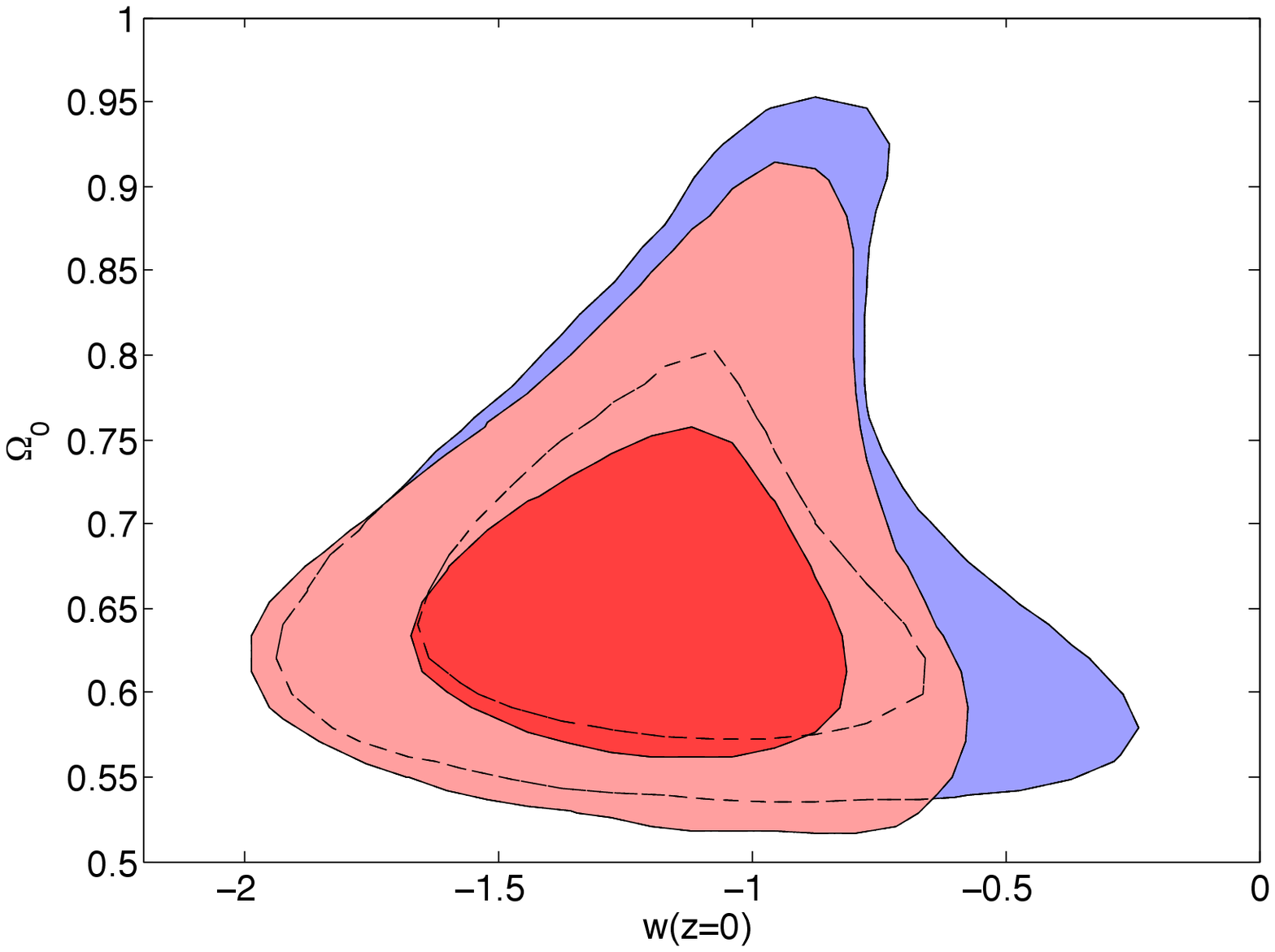} \\
	\end{tabular}
\vspace{-30mm}
	\caption{2D 68\% (dark shading) and 95\% (light shading) marginalised contours in the $\we|_{z=0}$--$\Omega_0$ plane. In both panels red regions correspond to the DE--clock and blue regions to CPL.  \textit{Left panel}: Flat priors on $w_0$ and $w_1$. Notice that $\we|_{z=0}$ is highly correlated with $\Omega_0$. \textit{Right panel}: Flat priors on $\we|_{z=0}$ and $\we^\prime|_{z=0}$. The correlation is reduced.}
	\label{fig:w0-Omega0planes}
\end{figure*}

With flat priors on $w_0$ and $w_1$ (so non--flat priors on $\we|_{z=0}$ and $\we^\prime|_{z=0}$), we find that $\we|_{z=0}$ in our DE--clock parametrisation is highly correlated with $\Omega_0$ as is shown in the left panel of Fig.~\ref{fig:w0-Omega0planes}.  The right panel of Fig.~\ref{fig:w0-Omega0planes} shows the same contours, but this time with flat priors on $\we|_{z=0}$ and $\we^\prime|_{z=0}$, the effect of which is to reduce the correlation between $\we|_{z=0}$ and $\we^\prime|_{z=0}$. 

The objective of this section has not been to distinguish which of the three parametrisations is best favoured by the data (which would require a full model comparison exercise), but rather to illustrate that our parametrisation is more sensitive to small deviations away from $\we=-1$, compared to CPL and GE.


\begin{figure*}[t]
	\begin{tabular}{ccc}
		\includegraphics[width=5.9cm]{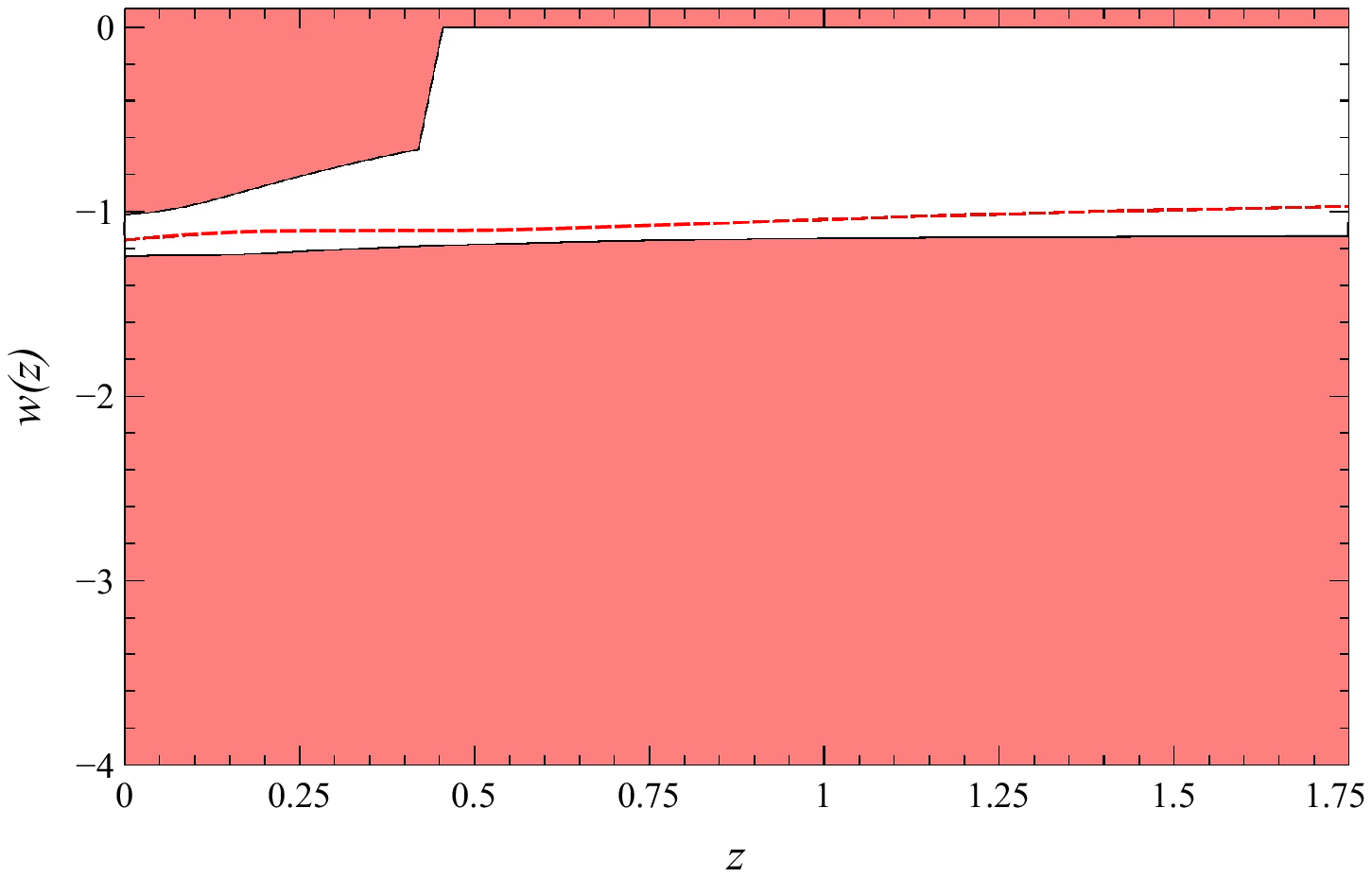} &	
		\includegraphics[width=5.9cm]{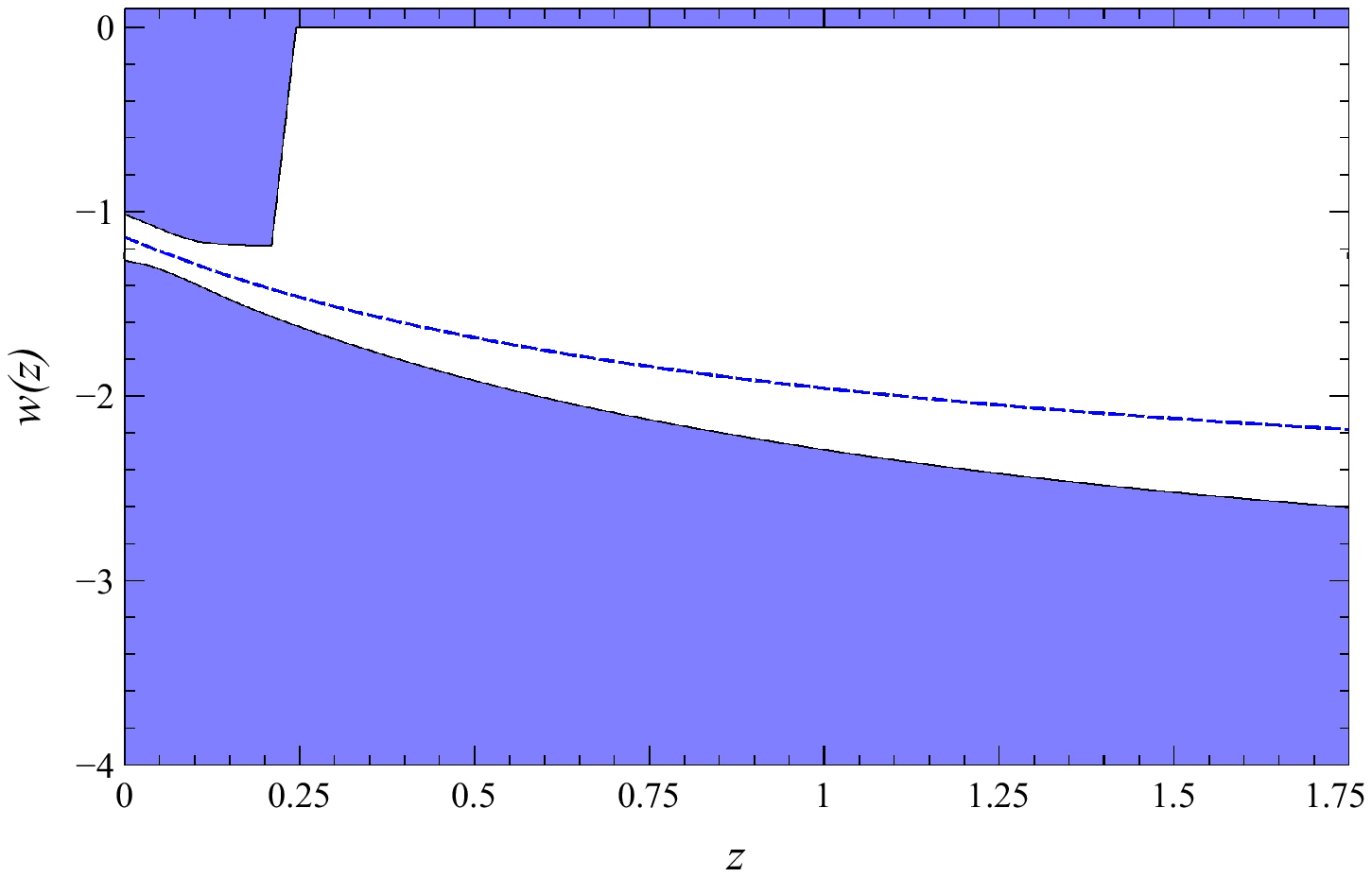} &
		\includegraphics[width=5.9cm]{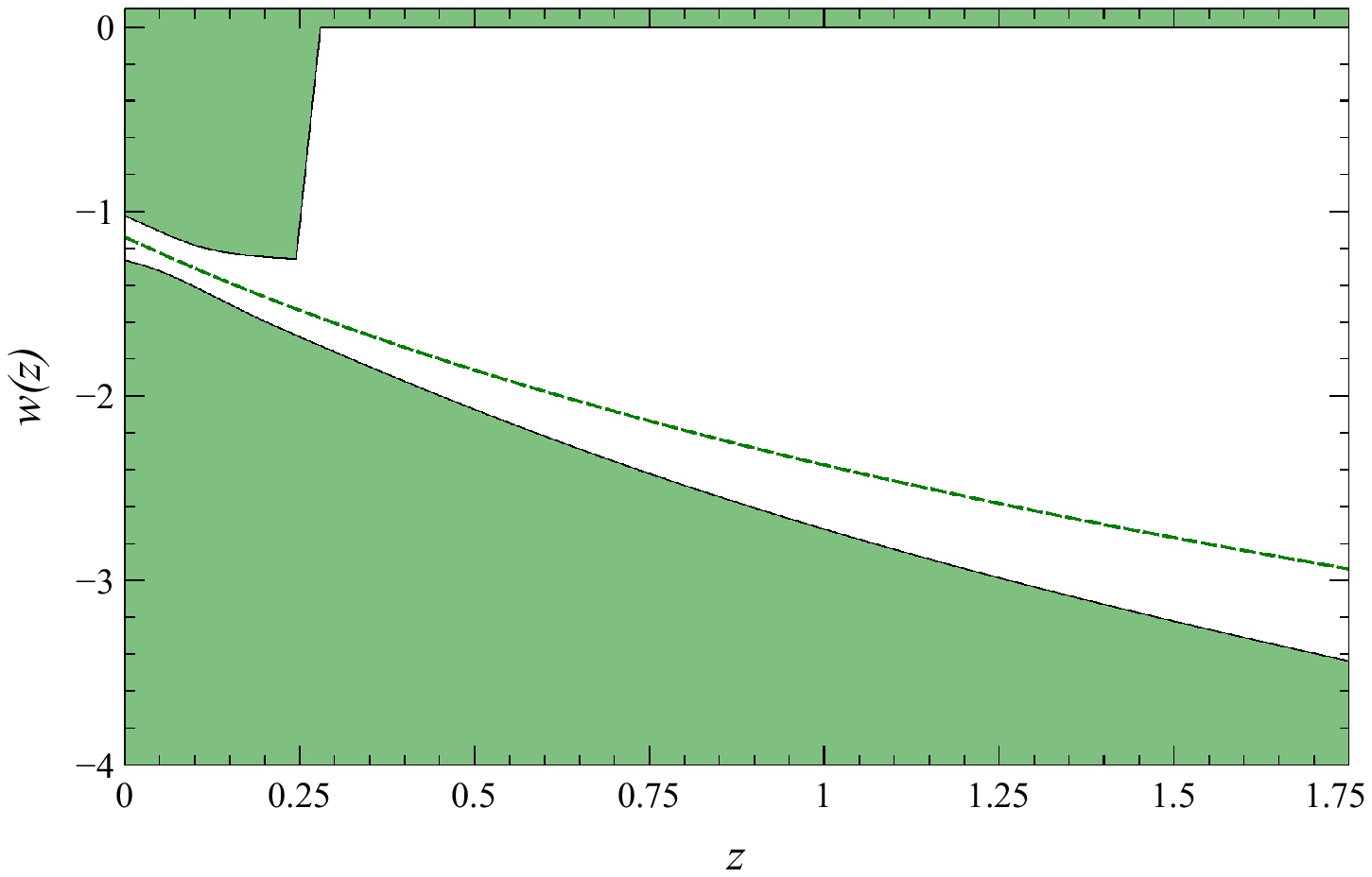}
	\end{tabular}
	\caption{The redshift evolution of the dark energy equation of state $\we$ for dark energy clock (red), CPL (blue) and GE (green) parametrisations. The shaded regions correspond to the values of $\we(z)$ that are ruled out at $68\%$ CL. Notice that the upper $68\%$ CL  limit on $\we$ quickly jumps to zero for all three parametrisations. This is because the hard prior $\we(z)<0$ cuts off the distribution for $\we(z)$ beyond a given redshift, and so the upper $68\%$ CL  limit on $\we$ at these redshifts is simply $\we<0$. Whilst the CPL and GE parametrisations allow $\we(z)$ to stray well below $-1$, our dark energy clock constrains $\we(z)$ to be close to $-1$ across the entire redshift range of interest.}
	\label{fig:evolve-w}
\end{figure*}

We plot in Fig.~\ref{fig:evolve-w} the redshift evolution of the dark energy equation of state for each parametrisation. The shaded regions correspond to the values of $\we(z)$ that are ruled out at $68\%$ CL, and were computed directly from the MCMC chains, and not through Gaussian error propagation. That is, for each sample $\sample$ in the chain we generate the distributions $\we(z)^{\rm CPL}$ (Eq.~(\ref{eq:CPL})) and $\we(z)^{\rm GE}$ (Eq.~(\ref{eq:GE})) over the redshift range where the data lie, $z=0 - 1.75$. From these distributions, one can easily compute the confidence regions. To generate the equivalent distributions using our dark energy clock parametrisation, $\we(\Oe)=w_0+w_1\Oe$, we take each sample $\sample$ and convert from redshift to $\Oe$ using the bisection algorithm described in Section.~\ref{sec:declock}. 

As can been seen from Fig.~\ref{fig:evolve-w}, the upper $68\%$ CL  limit on $\we$ quickly jumps to zero for all three parametrisations. This is because the hard prior $\we(z)<0$ cuts off the distribution for $\we(z)$ beyond a given redshift, and so the upper $68\%$ CL  limit on $\we$ at these redshifts is simply $\we<0$.  Whilst the CPL and GE parametrisations allow $\we(z)$ to stray well below $-1$, our dark energy clock constrains $\we(z)$ to be close to $-1$ across the entire redshift range of interest. This is due to two effects. The first is the lower value of $\we^\prime|_{z=0}$, (or more precisely $w_0$) for our parametrisation compared to CPL and GE. The second, more important effect is that our expansion parameter $\Oe(z)$ decays much faster with increasing redshift compared to the expansion parameters $z/(1+z)$ and ${\rm ln}\,[1/(1+z)]$ of the CPL and GE parametrisations respectively. For example, we find $\Oe(z=1)\approx0.05$, whilst $z/(1+z)=0.5$ and ${\rm ln}\,[1/(1+z)]=-0.69$ at $z=1$. Both of these effects keep $\we(z)$ close to $-1$.

%
%

\section{Comparing to Theory}\label{sec:theory}

As discussed in Section~\ref{sec:declock}, since we expand in a basis of orthogonal Chebyshev polynomials, any dark energy model which predicts a monotonic equation of state $w_{\rm theory}(\Oe)$, can be directly related to our parametrisation. We illustrate this by placing constraints on the Chebyshev expansion coefficients (the $\tw_n$ of Eq.~(\ref{eq:cheby2coeffs})) of two popular quintessence models, the single exponential potential (1EXP)~\cite{Ferreira:1997hj,Copeland:1997et} and the supergravity inspired potential (SUGRA)~\cite{Brax:1999gp}.  Using the $H_0$, $H(z)$ and SNIa data discussed in Section~\ref{sec:data}, we vary the base set of parameters 
\be
\sample_{\rm theory}  = \{ \Omega_mh^2,\, H_0,\,  \lambda\} \,,
\label{eq:quintBaseParams}
\ee
of both quintessence models, and use a modified version of the \texttt{CosmoMC} code~\cite{Lewis:2002ah} to sample from the joint posterior distribution of these parameters. For each sample $\sample_{\rm theory}$ in the MCMC chains, we can construct the equations of state, $w_{\rm 1EXP}(\Oe)$ and $w_{\rm SUGRA}(\Oe)$. Once these functions are known, the expansion coefficients, $\tw_n$, of each sample may be extracted by appealing to Eq.~(\ref{eq:cheby2coeffs}). If a sufficient number of samples are taken, we can generate distributions for the $\tw_n$.  The condition that $\Oe(t)$ must be a monotonic function of time, results in the prior $w_{\rm 1EXP}(\Oe)$ and $w_{\rm SUGRA}(\Oe)$ $<0$ over the redshift range of interest.  We choose the region over which the Chebyshev polynomials are orthogonal to be $\Oe=[\Oe(z=1.75),\,\Omega_0]$, which spans the redshift range of the data.

\begin{figure}[t]
\vspace{-30mm}
	\includegraphics[width=9.2cm]{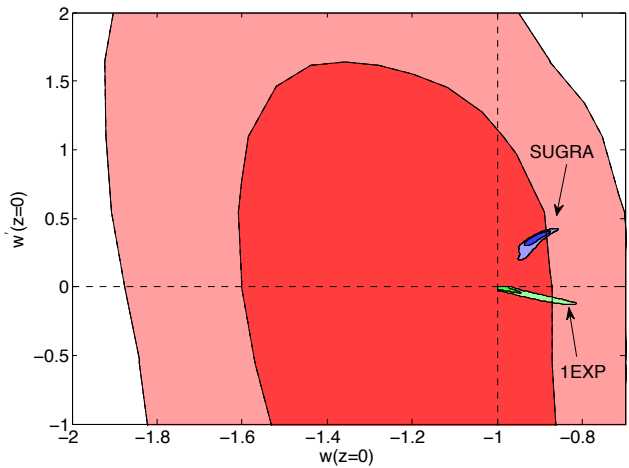}\\
\vspace{-30mm}
	\caption{The 2D 68\% (dark shading) and 95\% (light shading) marginalised contours in the $\we|_{z=0}$--$\we^\prime|_{z=0}$ plane for the 1EXP and SUGRA quintessence models superimposed upon the corresponding contours of the dark energy clock parametrisation.}
	\label{fig:TSPC-theory_w-wprime}
\end{figure}

In Table~\ref{tab:theory_coeffs} we quote the maximum likelihood values and the marginalised $68\%$ confidence limits of the Chebyshev expansion coefficients $\tw_n$ for the 1EXP and SUGRA models. We also give constraints on the derived parameters $\we|_{z=0}$ and $\we^\prime|_{z=0}$. We compare these constraints in Fig.~\ref{fig:TSPC-theory_w-wprime} by superimposing the 2D marginalised contours in the $\we|_{z=0}$--$\we^\prime|_{z=0}$ plane for the two quintessence models upon the contours of our dark energy cosmic clock.\\
\renewcommand*\arraystretch{1.9}
\begin{center}
\begin{table}[t]
\hfill{}
\begin{tabular}{c|c|c}
\hline
\hline
                                                      &  1EXP                                             	&      SUGRA  \\
\hline
    $\tw_0$  				&      $-0.975\,\, (-1.0\,,-0.972)$  	&      $-0.726\,\, (-0.756\,,-0.696)$   \\
    $\tw_1$  				&      $0.01\,\, (0.0\,,0.012)$    		&       $-0.093\,\, (-0.104\,,-0.092)$ \\
\hline
    $w_0$  					&      $-1.006\,\, (-1.008\,,-1.0)$  	&      $-0.298\,\, (-0.312\,,-0.20)$   \\
    $w_1$  					&      $0.069\,\, (0.0\,,0.080)$    		&       $-0.770\,\, (-0.90\,,-0.758)$ \\
\hline
   $\we|_{z=0}$  				&      $-0.954\,\, (-1.0\,,-0.947)$     	 &     $-0.912\,\, (-0.930\,,-0.893)$   \\
    $\we^\prime|_{z=0}$        	&      $-0.03\,\, (-0.04\,,0.0)$     	 	&     $0.339\,\, (0.298\,,0.379)$   \\
\end{tabular}
\hfill{}
\caption{Maximum likelihood values and marginalised $68\%$ confidence limits ($lower\,,upper$) for the first two Chebyshev expansion coefficients corresponding to the 1EXP and SUGRA quintessence models. We also quote constraints on the derived parameters $w_n$ (which are related to the $\tw_n$ by Eq.~(\ref{eq:wtowtilde})), and the equation of state today, $\we|_{z=0}$, and its derivative, $\we^\prime|_{z=0}$.}
\label{tab:theory_coeffs}
\end{table}
\end{center}
As can be seen from the figure, both quintessence models are consistent with the constraints on the dark energy equation of state at $68\%$ CL using our parametrisation. It is in this way that models of dark energy can be compared to our parametrisation, much like theoretical predictions of selected inflationary models superimposed upon the marginalized confidence regions for $n_s$ and $r_{0.002}$ using the recent Planck data~\cite{Ade:2013lta}.

We note that even though we have assumed flat priors on all base parameters, the priors on the expansion coefficients $\tw_n$ will not be flat. This is because the integral in Eq.~(\ref{eq:cheby2coeffs}) will in general be some complicated function of the base parameters. Furthermore, the limits $\Oe^{\rm min}=\Oe(z=1.75)$ and $\Oe^{\rm max}=\Omega_0$ of the integral Eq.~(\ref{eq:cheby2coeffs}) will themselves have probability distributions. A more careful and detailed  comparison involving identical priors will be left to future work, but we include this simpler study here in order to give the reader a general impression.

\section{Discussion and Conclusions}\label{sec:discussion}

Various astrophysical constraints suggest that the dark energy density $\Oe$ may have increased monotonically throughout cosmic history~\cite{Bean:2001wt,Reichardt:2011fv,Hinshaw:2012aka}. Acknowledging this apparent monotonicity we have introduced a new parametrisation of the dark energy equation of state which uses the dark energy density $\Oe(t)$ as a cosmic clock.  Our parametrisation has several advantages, perhaps the most important being that $\Oe$ is a physical quantity, directly related to the properties of dark energy. Furthermore, assuming $0<\Oe<1$, it makes for an ideal expansion parameter since $\Oe$ is a naturally small number. 
By fitting to SNIa and $H(z)$ data, we have demonstrated that constraints obtained on the expansion parameters of our parametrisation are tighter than the corresponding parameters of the popular Chevallier--Polarski--Linder (CPL) and Gerke and Efstathiou (GE) parametrisations. Furthermore, we have shown that our parametrisation is robust to the choice of prior. Expanding in orthogonal polynomials also allows us to relate models of dark energy directly to our parametrisation, which we have illustrated by placing constraints on the expansion coefficients extracted from two popular quintessence models.

The dark energy density $\Oe$ can only be used as a cosmic clock if it is a monotonic function of time. As can be seen from Eq.~(\ref{eq:Oedot1}), for a Universe containing dark energy, pressureless matter ($w_{\rm m}=0$), and radiation ($w_\g=\frac13$), there are turning points in $\Oe$ whenever $\we=\frac13$ in the radiation dominated era, and in the matter era whenever $\we=0$. Many scalar field models of dark energy possess scaling solutions on which the dark energy equation of state $\we$ `tracks' that of the dominant background component (see e.g.~\cite{Ferreira:1997hj,Copeland:1997et}). In such models, monotonicity of $\Oe$ would be spoiled as $\we$ evolves through the radiation and matter dominated eras and transitions towards $\we\sim-1$ today. In this paper, we have used our parametrisation to probe the dynamics of dark energy at low redshifts, $z=0-1.75$. It is presumably safe to assume that $\Oe$ remains monotonic over this redshift range, since to break monotonicity between $z=0$ and $z\sim2$ would require a rapidly varying $\we$, which is not favoured by existing analysis~\cite{Said:2013jxa,Sendra:2011pt,BenitezHerrera:2011wu}. Hence, in our fitting to data we imposed the hard prior $\we(z)<0$ for $z=0-1.75$.

To probe the high redshift behaviour of dark energy this prior would need to be removed, since it would be unrealistic to say with any sort of certainty that $\we(z)<0$ throughout the entire cosmic history. Hence, to accommodate high redshift data the non--monotonicity of $\Oe$ would need to be accounted for. This could be achieved by piece--wise parametrising $\we$ in regions of monotonic $\Oe$. For example, if radiation can be neglected, these regions are defined by the roots of the polynomial $\we(\Oe)=\sum^\infty_n \tw_n \tilde{U}_n(\Oe)=0$. If $\we(\Oe)=w_0+w_1\Oe$ and $w_1>0$, then there would be two distinct regions:  $\dot{\Omega}_{\rm e}>0$ for $\Oe<-w_0/w_1$ (I) and $\dot{\Omega}_{\rm e}<0$ for $\Oe>-w_0/w_1$ (II).   One would then write 
\be
\we(\Oe)=
 \begin{cases}
 w_0^{\rm (I)} + w_1^{\rm (I)}\Oe\,, &  \Oe<-\frac{w_0}{w_1} \\
 w_0^{\rm (II)} + w_1^{\rm (II)}\Oe\,, &  \Oe>-\frac{w_0}{w_1}
 \end{cases}
\ee
and so there would be four free parameters in total. To accommodate CMB data, radiation can not be neglected, but such regions of monotonicity can still be defined: one would need to compute the roots of Eq.~(\ref{eq:Oedot1}) exactly, which could be performed numerically. In the same fashion, non--zero cosmic curvature $\Omega_K\neq0$ can be easily accommodated: $\we$ can be piece--wise parametrised in regions of monotonic $\Oe$, where the boundaries of the distinct regions are again given by the roots of Eq.~(\ref{eq:Oedot1}), with $\Omega_K\neq0$.

Such `binning' of $\we$ into different regions is reminiscent of the principle component approach to constraining dark energy~\cite{Huterer:2002hy} (see also~\cite{Albrecht:2009ct}). However, division of $\we$ into regions of monotonic $\Oe$ would yield bins of non--constant width (the width would depend upon the sample $\sample$ -- see Eq.~(\ref{eq:baseParams})), unlike the constant redshift bin width, $\Delta z$ adopted in principle component analysis. Furthermore, across the finite width of each bin of monotonic $\Oe$, the equation of state would be free to vary, unlike the principle component approach, where for each redshift bin $z_i$, the value of $w_i$ in that bin is constant across its width $\Delta z$.

Finally, it is interesting to make the connection between our parametrisation and the dark energy `flow parameter'
\be
F\equiv\frac{1+\we}{\Oe\lambda^2}\,,
\ee
that was introduced in~\cite{Cahn:2008gk,Cortes:2009kc} (see also~\cite{Scherrer:2007pu}). Here, $\lambda=-V_\phi/V$, where $V(\phi)$ is the scalar field dark energy potential and $V_\phi$ is the derivative of $V$ with respect to $\phi$. By considering general dark energy models where the field either accelerates or decelerates down its potential toward its minimum (dubbed `thawing' or `freezing' field evolution~\cite{Caldwell:2005tm}), the authors of~\cite{Cahn:2008gk} were able to demonstrate that $F$ remains nearly conserved until quite recent times, $z\approx1-2$, after which dark energy finally begins to take over. This is despite $\we$, $\Oe$ and $\lambda$ all being dynamical. This constant nature of the flow parameter is a direct consequence of the fact that the dark energy field does not exist in a vacuum: instead it has been influenced by the long periods of radiation and matter dominated epochs prior to the current day.

If the parameter $\lambda$ is a constant, (as is the case for exponential potentials) or remains approximately constant, then we have $\we=-1+F\lambda^2\Oe$, which looks very much like our dark energy clock parametrisation with $w_0=-1$ and $w_1=F\lambda^2$. This indicates that, so long as $F$ remains approximately constant throughout the long radiation and matter dominated eras, then the behaviour of a wide range of scalar field dark models should be well captured by $\we(\Oe)=w_0+w_1\Oe$.


\section*{Acknowledgements}

The authors would like to thank Arman Shafieloo for inspiring discussions during the inception of this work. We would also like to thank Adam Moss, Mattia Fornasa, Anastasios Avgoustidis and especially Ren\'ee Hlozek for useful discussions. ERMT is supported by the University of Nottingham, EJC acknowledges The Royal Society, Leverhulme and STFC for financial support and AP and CS were funded by Royal Society University Research Fellowships. ERMT would like to thank Sir Bradley Wiggins for winning the 2012 Tour de France.

\appendix

\section{The Chebyshev Polynomials of the Second Kind}\label{appdx:ChebyPolys}

The Chebyshev polynomials of the second kind are defined by the recurrence relation
\bea
U_0(x)&=&1\,, \quad  U_1(x)=2x\,, \nonumber \\
U_{n+1}(x)&=&2xU_n(x)-U_{n-1}(x)\,,
\label{eq:cheby2recurrence}
\eea
and obey the following orthogonality condition
\be
\int^1_{-1}U_n(x)U_m(x)\sqrt{1-x^2}{\rm d}x=\frac{\pi}{2}\delta_{nm}\,.
\label{eq:cheby2ortho}
\ee
We can shift the interval over which the polynomials are orthogonal by choosing a suitable inner product. Let $\tilde{U}_n(x)=U_n(f(x))$ where $f(x)$ is monotonic on $[a,b]$, and satisfies $f(a)=-1$ and $f(b)=1$. The simplest choice is 
\be
f(x)=\frac{2x-a-b}{b-a}\,.
\label{eq:fx}
\ee
\renewcommand*\arraystretch{2.5}
\begin{center}
\begin{table*}[t]
\hfill{}
\begin{tabular}{c|c|c|c|c}
\hline
\hline
  			   &	$\alpha$                  &    $\beta$                   &                   $\g$                                   &   $F(\Oe)$ \\
\hline
$w_{0,1,2}\neq0$    &    $\frac{1+w_0}{w_0}$    &  $-\frac{1+w_T}{w_T}$    &   $-\frac{1}{2w_0}\frac{w_1+w_2}{w_T}$ & $ \frac{1}{w_0w_Tq} \left[w_1(w_1+w_2)-2w_0w_2\right]\,{\rm arctan}\left[\frac{2w_2q(\Omega_0-\Oe)}{(2w_2\Omega_0+w_1)(2w_2\Oe+w_1)+q^2}\right]$ \\
\hline
$w_0=0\,,\;w_{1,2}\neq0$    &    $1+\frac{w_1-w_2}{w_1^2}$    &  $-\frac{1+w_T}{w_T}$    &   $\frac{w_2^2}{w_1^2}\frac{1}{w_T}$ & $ \frac{1}{w_1}\left[\frac{1}{\Omega_0} -\frac{1}{\Oe}\right]$ \\
\hline
$w_1=0\,,\;w_{0,2}\neq0$    &    $\frac{1+w_0}{w_0}$    &  $-\frac{1+w_T}{w_T}$    &   $-\frac{w_2}{2w_0w_T}$ & $\frac{w_2}{w_T\sqrt{w_0w_2}}\,{\rm arctan}\left[\frac{\sqrt{\frac{w_2}{w_0}}(\Oe-\Omega_0)}{1+\frac{w_2}{w_0}\Oe\Omega_0} \right]$ \\
\hline
$w_2=0\,,\;w_{0,1}\neq0$    &    $\frac{1+w_0}{w_0}$    &  $-\frac{1+w_T}{w_T}$    &   $-\frac{w_1}{w_0w_T}$      &     $0 $ \\
\hline
$w_{0,1}=0\,,\;w_2\neq0$    &    $\frac{1+w_2}{w_2}$    &  $-\frac{1+w_T}{w_T}$    &   $0$    &   $  \frac{1}{w_2}\left[\frac{1}{\Omega_0}-\frac{1}{\Oe}+\frac{1}{2\Omega_0^2}-\frac{1}{2\Oe^2}  \right]$ \\
\hline
$w_{0,2}=0\,,\;w_1\neq0$    &    $\frac{1+w_1}{w_1}$    &  $-\frac{1+w_T}{w_T}$    &   $0$    &   $  \frac{1}{w_1}\left[ \frac{1}{\Omega_0} -\frac{1}{\Oe}\right]$ \\
\hline
$w_{1,2}=0\,,\;w_0\neq0$    &    $\frac{1+w_0}{w_0}$    &  $-\frac{1+w_T}{w_T}$    &   $0$    &   $  0$ \\
\hline
\end{tabular}
\hfill{}
\caption{The values of the powers $\alpha$, $\beta$ and $\g$ and the function $F(\Oe)$ for particular cases of zero $w_0$, $w_1$ or $w_2$.  Here, $q=\sqrt{4w_0w_2-w_1^2}$, and $w_T=w_0+w_1+w_2$. Notice that $F(\Oe)$ may always be kept real by using the relation ${\rm arctan}(iz)=i\,{\rm arctanh}(z)$ when necessary (for example if $4w_0w_2<w_1^2$ in the definition of $q$). }
\label{tab:particularsols}
\end{table*}
\end{center}
Then, the orthogonality condition becomes 
\be
 \int_{a}^{b}  \tilde{U}_n(x)\tilde{U}_m(x)\sqrt{(x-a)(b-x)}{\rm d}x=\frac{\pi}{8}(b-a)^2\delta_{nm}\,.
\label{eq:cheby2orthoab}
\ee
Any function $g(x)$ which is continuous in the interval of orthogonality $[a,b]$, may be expanded as a series of Chebyshev polynomials:
\be
g(x)=\sum_n^\infty \tw_n \tilde{U}_n(x)\,,
\label{eq:chebyExpandg}
\ee
From the orthogonality condition, Eq.~(\ref{eq:cheby2orthoab}) we have:
\be
\tilde{w}_n=\frac{8}{\pi(b-a)^2} \int_{a}^{b}g(x) \tilde{U}_n(x) \sqrt{(x-a)(b-x)}\, {\rm d}x  \,.
\label{eq:cheby2coeffs1}
\ee
Under certain conditions of the interpolated function $g(x)$ (Dini–-Lipschitz continuity), the Chebyshev interpolation converges when the number of nodes tends to infinity. For numerical implementation, it is convenient to make the change of variable $x = a+(b-a)\,{\rm sin}^2\theta$ giving
\bea
\tilde{w}_n&=&\frac{16}{\pi} \int_{0}^{\pi/2} \Big[g \left( a+(b-a)\,{\rm sin}^2\theta  \right) \nonumber \\ 
							&\times& \tilde{U}_n\left( a+(b-a)\,{\rm sin}^2\theta  \right) {\rm sin}^2\theta\, {\rm cos}^2\theta\Big]\,{\rm d}\theta\,. \nonumber
\label{eq:cheb2_coeffs2}
\eea
In the case where $g(x)=\we(\Oe)$, and $b=\Oe^{\rm max}$, $a=\Oe^{\rm min}$, we can write Eq.~(\ref{eq:chebyExpandg}) at second order as $\we(\Oe)=w_0+w_1\Oe+w_2\Oe^2$. The $w_n$ are given in terms of the $\tw_n$ as:
\begin{eqnarray}
w_0 &=& \tw_0 -\frac{2(\Oemin+\Oemax)}{\Oemax-\Oemin}\tw_1 \nonumber \\
	&+&\left[\frac{4(\Oemin+\Oemax)^2}{(\Oemax-\Oemin)^2}-1\right]\tw_2 \,,  \nonumber \\
w_1 &=& \frac{4}{\Oemax-\Oemin}\tw_1 - \frac{16(\Oemin+\Oemax)}{(\Oemax-\Oemin)^2}\tw_2\,, \nonumber \\
w_2 &=& \frac{16}{(\Oemax-\Oemin)^2}\tw_2 \,.
\label{eq:wtowtilde}
\end{eqnarray}
%

\section{The $\Oe$--clock particular solutions}\label{appdx:particularsols}

The powers $\alpha$, $\beta$ and $\g$ that appear in the solutions for $a(\Oe)$ and $\rhoe(\Oe)$ (Eqs.~(\ref{eq:scalefactorsol}) and (\ref{eq:rhoesol})) are combinations of $w_0$, $w_1$ and $w_2$, and correspond to coefficients of the partial fraction expansions of Eqs.~(\ref{eq:scalefactorintegral}) and (\ref{eq:rhoeintegral}). As a result, if any one or more of the coefficients $w_0$, $w_1$ or $w_2$ is exactly zero, then the powers $\alpha$, $\beta$ and $\g$ and the function $F(\Oe)$ will change. In table~\ref{tab:particularsols} we list all the possible solutions for different combinations of $w_0$, $w_1$ and $w_2$, where one or more of them is zero.


\bibliographystyle{apsrev4-1}
\bibliography{bib_file}

\end{document}